**Ramp Josephson Junctions of Al/Ti/$Sr_2RuO_4$: Observation of Single-Domain Quantum Oscillations and the Detection of Chiral Surface Current**

Zixuan Li[1,∀], Yiqun Alex Ying[1,#], Brian M. Zakrzewski[1,*], Yan Xin[2], Yu Wang[1], Zhiqiang Mao[1], and Ying Liu[1,†]

[1]Department of Physics and Materials Research Institute, the Pennsylvania State University, University Park, PA 16802, U.S.A.

[2]National High Magnetic Field Laboratory, Florida State University, Tallahassee, Florida 32310, U.S.A.

[∀,#,*]These authors contributed to this work equally.

[†]Email: yxl15@psu.edu.

## Abstract

The determination of how the phase of the superconducting order parameter in a superconductor varies with the spatial direction, which can be done only through the Josephson-effect-based phase-sensitive measurements, is crucial for the establishment of the precise pairing symmetry of the superconductor. So far, such measurements have been done on high-$T_c$ cuprate superconductors but only at a couple of directions for $Sr_2RuO_4$ because of the difficulty in preparing Josephson junctions between $Sr_2RuO_4$ and an *s*-wave superconductor with a chosen mutual orientation. Another long-standing issue in $Sr_2RuO_4$, which was shown previously to feature a spontaneously broken time-reversal symmetry by muon spin rotation and other measurements, is that the expected presence of chiral surface currents, domains, and domain walls is yet to be explicitly shown experimentally. To address these issues, we have long sought the preparation of high-quality Josephson junctions between $Sr_2RuO_4$ and a conventional *s*-wave with a controllable orientation relative to symmetry axes in $Sr_2RuO_4$. We report in this article the successful fabrication of ramp Josephson junctions of Al/Ti/$Sr_2RuO_4$ on thin single crystals of $Sr_2RuO_4$ obtained by mechanical exfoliation. These junctions were found to show high-quality quantum oscillations consistent with a single-domain Josephson coupling. The normal junction resistance was found to depend extremely sensitively on the supercurrent flowing in the $Sr_2RuO_4$ crystal on which the Josephson junction was made. This finding was used in the present work to provide an estimate of the size of the chiral surface current, which is shown to agree with its upper bound established previously.

The symmetry properties of the superconducting order parameter (OP) are among the most important characteristics of unconventional superconductors. According to the Volovik-Gor'kov theory of their classification[1], the superconducting OP is given by the basis functions of the irreducible representations (irreps) of the group characterizing the normal-state symmetry of the superconductor, G x T x U(1), where G is the point group characterizing the crystalline symmetry, T is the time-reversal symmetry (TRS), and U(1) is the gauge symmetry, respectively. The spontaneous breaking of symmetries beyond the U(1) symmetry makes the superconductor unconventional. If the TRS broken state of the superconductor is that of the spin-triplet, chiral *p*-wave in the two dimensions (2D), Majorana zero-energy modes (MZEMs) are expected[2], which are believed to be useful for realizing the fault-tolerant topological quantum computing[3,4,5].

$Sr_2RuO_4$, which features a crystal structure of the point group $D_{4h}$ and space group I4/mmm (No. 139)[6,7], is isostructural with the single-layer high-$T_c$ cuprates. The normal-state symmetry of $Sr_2RuO_4$ allows five even- and five odd-parity pairing states based on the irreps of the $D_{4h}$ x T x U(1) group. Because of the presence of inversion symmetry, the symmetry properties of the orbital and spin parts of the superconducting OP are correlated, with the even parity being spin singlet and the odd parity being spin triplet. $Sr_2RuO_4$ was predicted to feature odd-parity, spin-triplet pairing[8,9] shortly after superconductivity was discovered in this material[10]. Among the odd-parity, spin-triplet pairing states, the superconducting OP of the $\Gamma_5^-$ can be specified by a d-vector, $\vec{d} = \hat{z}(k_x \pm ik_y)$, where $\vec{k} = (k_x, k_y)$ is the wavevector of the Bloch function, known as the chiral *p*-wave pairing state. Here the amplitude of the d-vector is the superconducting energy gap and its direction is the one onto which the projection of the spin of the Cooper pair is zero. The two-dimensional representation of the $\Gamma_5^-$ state implies that the chiral surface currents as well as domains (the $k_x + ik_y$ and $k_x - ik_y$ domains) and domain walls will be present in the state[11,12]. Matsumoto and Sigrist investigated these properties in the zero-temperature limit for $Sr_2RuO_4$ using the quasi-classical Green's function formalism[13].

The symmetry properties of the orbital part of the OP are probed by the phase-sensitive measurements[14] while those of the spin part by the spin-susceptibility measurements[15]. In the former, the angle-dependent Josephson coupling between an *s*-wave and unconventional superconductor, facilitated by a finite spin-orbital coupling[16,17], is used to determine the angle dependence of the phase of the superconducting OP. Class I phase-sensitive experiments, in which the selection rule of the Josephson coupling between an *s*-wave superconductor and $Sr_2RuO_4$ in a single planar junction, are used to exclude symmetry-allowed pairing states. For $Sr_2RuO_4$, all even-parity, spin-singlet as well as all odd-parity, spin-triplet states *except* the chiral *p*-wave state are excluded[18]. In particular, even though the spin-singlet, chiral *d*-wave state ($\Gamma_5^+$) will also allow a finite Josephson coupling with an *s*-wave superconductor if the junction plane is tilted away from being parallel with the *c* axis[19], the theoretically calculated Josephson coupling strength in the units of that given by the Ambegaokar-Baratoff value for an all-*s*-wave Josephson junction[20] is two to three orders of magnitude smaller than those obtained experimentally[18], making it unlikely that the chiral *d*-wave will be adopted by $Sr_2RuO_4$. In Class II phase-sensitive measurements based on superconducting quantum interference device (SQUID) consisting of two oppositely faced Josephson junctions (with 180-degree difference in their orientations) between an *s*-wave superconductor and $Sr_2RuO_4$ showed that the phase of the OP in $Sr_2RuO_4$ changes by π after the 180-degree rotation[21], provided the strongest evidence available so far that $Sr_2RuO_4$ is indeed an odd-parity, and therefore, spin-triplet superconductor.

The symmetry properties of the spin part of the OP are probed primarily by the behavior of the spin susceptibility through the nuclear magnetic resonance (NMR) Knight shift or polarized neutron scattering (PNS) measurements. The early Knight shift measurements on $Sr_2RuO_4$ showed a constant Knight shift across $T_c$[22], which however was found to be wrong due to the unwanted sample heating[23,24]. The new data revealed that the Knight shift normalized by its normal-state value decreases significantly as the temperature is lowered. At 25 mK and in the presence of an applied in-plane magnetic field of 0.28 T, the Knight shift was found to be roughly (12 – 20)% of the normal-state value (with large error bars)[25]. This led to the ongoing debate as to whether $Sr_2RuO_4$ is even a spin triplet. On the other hand, the Knight shift measurements cited above and that from the PNS measurements are inconsistent with one another beyond the error bars at the overlapping magnetic fields and comparable temperatures as shown in Figs. 1b and c of Ref. 26. Specifically, at 0.5 T, the PNS measurements showed a spin susceptibility value of 60% to 70% ($\pm$ 10%) of the normal-state value while the Knight shift measurements yielded roughly 35% ($\pm$ 10%) of the normal-state value. The spin susceptibility of the $\Gamma_5^-$ is generally believed to be constant across $T_c$, making $\Gamma_5^-$ unlikely to be adopted by $Sr_2RuO_4$. On the other hand, the $\Gamma_{1-4}^-$ states (the spin-triplet helical states with the d-vector in the *ab* plane) are not excluded for $Sr_2RuO_4$ if the Fermi liquid corrections are considered[27] since the spin susceptibility is expected to be reduced by the Fermi liquid parameter $F_{a0}$[28]. This parameter characterizes the anisotropic part of the interaction between two quasiparticles in the Fermi liquid due to the "molecular field" of the magnetic fluctuation[29], which is related to the Wilson ratio, $R_W$. For superfluid $^3$He, $R_W \approx 4$[30] , implying that[29] $F_{a0} \approx 1/R_w - 1 \approx -0.75$ (comparable with the experimental value of -0.7[31]) with the experimentally observed spin susceptibility of superfluid $^3$He[31] agreeing with the theoretical expectations[27]. For $Sr_2RuO_4$, $R_W \approx 1.8$[32], which should give make $F_{a0} \approx - 0.44$. As a result, the Fermi liquid corrections should lead to a spin susceptibility of the helical states in the zero-temperature limit roughly 35% of the normal state value, which seems to be consistent with the PNS but barely with Knight shift experiments even when the error bars are considered, an issue to be fully resolved[28].

Spontaneous breaking of TRS in $Sr_2RuO_4$ is nevertheless supported solidly by several experiments. In the muon spin relaxation (μSR) measurements, the zero-field relaxation rate of positive charged muons suddenly increased at $T_c$ and continues to increase as temperature decreases. Consistent results were obtained across measurements performed by several independent groups[33,34,35,36,37,38]. The polar Kerr effect measurements[39] on $Sr_2RuO_4$ also revealed a signal of spontaneous local magnetic field at the onset of superconductivity. In the phase sensitive experiments, the behavior of the SQUIDs consisting of two oppositely faced Josephson junctions between an *s*-wave superconductor and $Sr_2RuO_4$[21] is only possible if one of the two symmetry-allowed chiral states ($\Gamma_5^-$ or $\Gamma_5^+$) featuring the TRS breaking is adopted. But $\Gamma_5^+$ is excluded as discussed above. The TRS breaking will also lead to consequences in quantum oscillations due to the formation of domains and domain walls, which were also observed[40]. Early attempts to detect the chiral surface currents using scanning SQUIDs[41,42], however, yielded only an upper bound for the magnetic field produced by the chiral surface currents, which is consistent with the Hall probe measurements[43] possessing a sensitivity lower than that of the scanning SQUIDs. A recent experiment sought evidence for TRS breaking through the detection of the nonreciprocal Josephson effect in $Sr_2RuO_4$/Nb junctions[44]. The authors argued that the breaking of the TRS in $Sr_2RuO_4$ should lead to nonreciprocal behavior in the critical current of the Josephson junction with respect to the simultaneous reversals of the directions of the current and the magnetic field, which is correct in principle, and concluded that the TRS is not broken in $Sr_2RuO_4$ since no nonreciprocal behavior was found in their data. However, this argument misses the important point that the quantitative size of the effect is directly related to the magnetic field generated by the chiral surface currents[45]. While the TRS breaking will indeed lead to the nonreciprocal behavior, it is observable only when the chiral surface current induced magnetic flux is sufficiently large. The upper limit established by scanning SQUID measurements[41,42] suggests that signal of nonreciprocal behavior will be well below the detection limit of this experiment[44].

We report in this article the successful fabrication of ramp Josephson junctions of Al/Ti/$Sr_2RuO_4$ on thin single crystals of $Sr_2RuO_4$ and high-quality quantum oscillations as a function of magnetic field applied along the junction plane. The normal-state junction resistance was found, unexpectedly, to depend extremely sensitively on the supercurrent flowing in the $Sr_2RuO_4$ crystal on which the Josephson junction was made and used to provide an estimate of the size of the chiral surface current. The experimental design of this new detection scheme of the chiral surface current is shown schematically in Fig. 1a. A ramp junction prepared on a thin crystal of $Sr_2RuO_4$ forms a Josephson coupling between an *s*-wave superconductor (Al) and $Sr_2RuO_4$, featuring the chiral surface current, $J_{\mathrm{chi}}$, with a thin interface layer of Ti in between. The magnetic flux generated by the chiral surface current induces a Meissner current towards the interior of the crystal, $J_{\mathrm{Mei}}$, which flows in the opposite direction of the chiral surface current. The magnetic field produced by the chiral surface and Meissner currents, $\boldsymbol{H}_{\mathrm{E}}$, perpendicular to the currents and the *ab* plane of $Sr_2RuO_4$, is confined between the two currents. The magnetic field, $\boldsymbol{H}$, was applied along the junction plane and the *ab* plane in the current experiment. As a result, $\boldsymbol{H}$ and $\boldsymbol{H}_{\mathrm{E}}$ are perpendicular to one another with a magnitude of the total magnetic field, $\boldsymbol{H}_{\mathrm{T}}$, given by $H_{\mathrm{T}} = (H^2 + H_{\mathrm{E}}^2)^{1/2}$. The relevant area used to calculate the magnetic flux produced by $\boldsymbol{H}$ is given by $a_1 = L\,(\lambda_{\mathrm{Al}} + \lambda_{214,c} + d)$, where L is the length of the junction along the ramp, $\lambda_{\mathrm{Al}}$ is the penetration depth of Al film and $\lambda_{214,c}$ is that of $Sr_2RuO_4$ along the *c* axis when an in-plane field is applied, and d is the thickness of the non-superconducting layer of $Sr_2RuO_4$ and that of an interface layer (see below), which is expected to be two or three orders of magnitude smaller than $\lambda_{\mathrm{Al}} + \lambda_{214,c}$. However, to calculate the flux produced by $\boldsymbol{H}_{\mathrm{E}}$, a different loop must be used, resulting in an area of $a_2 = L\,(\lambda_{\mathrm{Al}} + \lambda_{214,ab} + d)$, where $\lambda_{214,ab}$ ($\ll \lambda_{214,c}$) is the penetration depth of $Sr_2RuO_4$ along the *ab* plane when a *c*-axis field is applied. The thickness of the normal layer, which is due to the surface of $Sr_2RuO_4$ being non-superconducting because of the extreme sensitivity of its superconductivity to disorder[46,47], is estimated to be on the order of several nm. The critical current and normal-state resistance of the junction, $I_c$ and $R_J$, are expected to oscillate with ***H***, with the period in the units of the applied magnetic field deduced from the magnetic flux quantum, $\Phi_0$ (= h/2e).

For the current experiment, ~ 1 μm thick crystal plates (flakes) with a lateral dimension of ~20 μm are obtained by mechanical exfoliation of a bulk crystal that is grown by the floating zone method specifically to be slightly Ru deficient, resulting in a $T_c$ lower than optimal. Selected crystal plates were transferred onto a clean $Si/SiO_2$ substrate with the *c* axis of the crystal perpendicular to the substrate. Around 200 nm thick $SiO_2$ was deposited on top of the plate as a protection layer prior to using a focused ion beam (FIB) of 30 kV Ga ions to cut a ramp out of the crystal. A subsequent 20-minute ion mill of 300 V Ar ions was used to clean up the re-deposition and the top layer of the crystal surface of the ramp which was implanted by Ga ions after unneeded parts of the crystal were removed. Electrical leads were defined by standard contact photolithography. After the development, a 30 second oxygen plasma operated at 100 mTorr and 100 W and a 5 minute *in-situ* ion mill of 100 V Ar ions to remove residue photoresist and adsorbed organics before metallization using the e-beam deposition of the 5 nm Ti and 200 nm Al. E-beam evaporated Al is chosen to be the *s*-wave superconductor for the ramp Josephson junction in the current experiment due to its long superconducting coherence. The Ti interlayer serves two purposes. First, it improves the adhesion of the Al film to the $Sr_2RuO_4$ surface and helps to prevent oxygen in $Sr_2RuO_4$ being extracted to form an $AlO_x$ layer, which was found to be too high a tunnel barrier to allow a decent Josephson coupling between Al and $Sr_2RuO_4$. This layer also helps improve the adhesion of Al on $Sr_2RuO_4$. Scanning electron microscopy (SEM) images of the device are shown in Figs. 1b-c. The ramp orientation with respect to the *a* or *b* axis is not determined in the current experiment even though it can be done using, for example, Raman scattering spectroscopy with polarized light[48]. As shown by transmission electron microscopy studies (Fig. 1d), with the help of the thin interlayer of Ti excellent contact between Al and $Sr_2RuO_4$ was achieved. We performed low temperature *d.c.* measurements in a $^3$He refrigerator with a base temperature of 0.38 K. All leads entering the cryostat were shielded and filtered by low-pass RC filters with a 3dB cut-off frequency around 600 kHz.

The junction and the crystal resistances ($R_J$ and $R_{cry}$) as a function of temperature ($T$) for Sample A were shown in Figs. 1e and 2a, respectively. The superconducting transition temperature of the crystal ($T_{c,cry}$) is lower than the optimal because it was grown Ru-deficient deliberately to allow the mechanical exfoliation of the crystal. The critical current for Sample A was found to be 280 μA at $T = 0.38$ K (Fig. 2b). Josephson coupling was obtained in two Al/Ti/$Sr_2RuO_4$ junction as seen in the *I-V* curves (Figs. 2c and d). As $H$ was varied, $I_c$ was found to show quantum oscillations (Figs. 2e and f) with a quality of oscillations never obtained on any junction involving $Sr_2RuO_4$ prior to the current work. For Sample A, the primary peak was found to fit better to the Airy than Fraunhofer pattern (see Supplementary Materials – SM), suggesting that the Josephson junction is reasonably uniform and furthermore of a circular (as opposed to rectangular) shape. The magnetic fields corresponding to a flux quantum were found to be 17 and 44 G for Junctions A and B, respectively. Using values of penetration depths, $\lambda^{Al} \approx 0.1$ μm (for Al films), $\lambda^{Sr2RuO4}_{H//ab} = 3.7$ μm, and d ≈ 10 nm, we can get the estimation dimension along the ramp for Junction A at 0.32 μm and Junction B at 0.12 μm, smaller than the width of the ramp (1.6 μm and 0.8 μm) as expected. It was found previously that the presence of domains leads to irregular patterns in quantum interference[40], which raises the question on whether the Josephson coupling between Al and $Sr_2RuO_4$ encompasses more than one domain in the current samples. In this regard, the Matsumoto and Sigrist calculations[13] revealed a domain wall width of 5 or 6 times of the in-plane superconducting coherence length, which is 67 nm for $Sr_2RuO_4$, comparable with the size of the Josephson junction. The above consideration together with the primary peak at zero magnetic flux and the fitting of quantum oscillation pattens strongly suggested that the junction used in the current study hosts a single ($k_x + ik_y$ or $k_x - ik_y$) domain.

The demonstration of the single-domain Josephson coupling in ramp junctions between an *s*-wave superconductor and $Sr_2RuO_4$ prepared by conventional nanofabrication techniques will help lay the foundation for the determination of the precise symmetry properties of the superconducting OP in $Sr_2RuO_4$ by phase-sensitive measurements. For the high-$T_c$ superconductors, the d-wave pairing symmetry was established largely by the phase-sensitive experiments[49,50]. Its precise pairing symmetry, however, was determined by the measurement of the precise angle dependence of the phase of the OP using the Geshkenbein-Larkin-Barone SQUIDs containing two Josephson junctions between Nb and $YBa_2Cu_3O_{7-x}$ with a precisely controlled relative orientation between the two junctions (respected to the crystal axes of $YBa_2Cu_3O_{7-x}$)[51]. The change in the spontaneously formed magnetic flux trapped in the SQUID leads to not only the confirmation of the predominant *d*-wave pairing in $YBa_2Cu_3O_{7-x}$ but also the discovery of an *s*-wave component in the superconducting OP roughly 5% of the d-wave component. The mixed pairing state in $YBa_2Cu_3O_{7-x}$ is consistent with the small orthorhombicity in the crystal structure making the point group symmetry of $YBa_2Cu_3O_{7-x}$ deviate slightly from the tetragonal crystalline symmetry of $D_{4h}$. The same phase-sensitive experiment can be performed on $Sr_2RuO_4$ given the growth of superconducting films of $Sr_2RuO_4$ demonstrated previously[52,53,54,55,56] and the successful preparation of single domain ramp junctions of Al/Ti/$Sr_2RuO_4$ shown in the present work.

Data shown in Fig. 2 have also revealed the sensitivity of $I_c$ with respect to the magnetic field applied along the junction plane, which is relevant to the detection of the magnetic field produced by the chiral surface and Meissner currents. However, their upper bound set previously[41,42], on the order of tens of mG, suggests that the change in $I_c$ caused by the magnetic field generated these currents will not be observable for the junction size achieved in the current work. The preparation of a Josephson junction with a sufficiently large area to detect the magnetic flux generated by a few tens of mG is not practical. Interestingly, however, $R_J$ measured with a current slightly larger than $I_c$ was found to be very sensitive to $I_{cry}$, the supercurrent flowing in the $Sr_2RuO_4$ crystal on which the junction was prepared (Fig. 1a). In fact, as will be seen below, an $I_{cry}$ on the order of 1 μA leads to a detectable change in $R_J$. Empirically the normal resistance of the Josephson junction seems to always depend on its critical current monotonically. As a result, it is natural to assume that as the magnetic flux enclosed in the junction varies, the junction resistance varies accordingly. On the other hand, using the real dimensions of Sample A, one can calculate in the London approximation[57] the magnetic field generated by an $I_{cry} \sim 1$ μA, which is on the order of 1 mG (see SM). It is unlikely that the sensitive dependence of $R_J$ on $I_{cry}$ observed here can be attributed to the dependence of $R_J$ on the tiny magnetic flux generated by $I_{cry}$.

The surface of the $Sr_2RuO_4$ crystal on which the junction is prepared is subject to three currents, $I$, $I_{cry}$ and $I_{chi}$. Even though $I$ is perpendicular to $I_{cry}$ and $I_{chi}$, the amplitude of the combined current could be sufficiently large to break Cooper pairs via the Volovik effect[58]. Experimentally, the superconducting energy gap in $Sr_2RuO_4$ was found to possess a deep minimum even though $\Gamma_5^-$ is fully gapped on any of its three-sheet cylindrical Fermi surface due to the established band-dependent[59] or odd-frequency[60] superconductivity in $Sr_2RuO_4$. In fact, small angle neutron scattering measurements[61] indicate that the minimal gap value is only about 15% of the maximal gap value. For our crystals with a lower than optimal $T_c$ of 1.1K, this small gap would behave effectively as a nodal gap down to even the base temperature, 0.38 K, used in the current experiment, which is consistent with the results of a recent μSR study[62]. In this picture, the sensitive of $R_J$ *vs.* $I_{cry}$ seen in the present work could be the result of the suppression of the superconducting energy gap due to the Volovik effect, which increases $R_J$. Another possibility is the phase slips caused by the crossing of the Josephson vortices driven by the total surface current. It should be noted that the tunnel barrier in our Josephson junctions of Al/Ti/$Sr_2RuO_4$ is naturally formed, making the Josephson vortices closer to the Abrikosov vortices featuring a normal core. As shown in Fig. 3a, when the magnetic flux line trapped in a Josephson vortex is along the *c* axis, the size of the vortex in the *ab* plane will roughly $\lambda^{\mathrm{Sr2RuO4}}_{\mathrm{H//c}} = 0.18$ μm, which will be slightly smaller than the range over which $J_{chi}$ is distributed (see below) according to the Matsumoto and Sigrist calculation[13]. As a result, $J_{chi}$ will apply a force along the horizontal direction (note that the Josephson vortex will not extend to the region of $J_{Mei}$, which suggests that the force from $J_{chi}$ will not be balanced out by $J_{Mei}$). This force will in turn possess a component along the junction plane (see Fig. 3a). Together with the measurement current, the chiral surface current will push the Josephson vortex to move, resulting in phase slips and an induced voltage according to the Josephson relation[63]. Interestingly, in both pictures described above, $R_J$ will be a minimum when the total current flowing on the surface is minimized at a $I_{cry} = I_{0,cry}$.

Data from detailed $R_J$ *vs.* $I_{cry}$ measurements showed that $R_J$ can be described by a parabolic function of $I_{cry}$, $R_J = R_J(0) + b\,(I_{cry} - I_{0,cry})^2$, where $b$ and $I_{0,cry}$ are constants. We performed multiple measurements on Junction A at a fixed temperature in three separate cooling downs in which the residue magnetic field was minimized. From the obtained set of offset values of $I_{0,cry}$ (see the histogram in Fig. 3c) the average $I_{0,cry}$ was roughly found to be 0.6 μA. To compare this value with the upper bound in the magnetic flux produced by the chiral surface and Meissner currents obtained in the scanning SQUID measurements[41,42] quantitatively, the distribution of current density in the crystal is calculated in the London approximation with the magnetic field produced by the current distribution calculated by the Biot-Savart law (for details see SM). These results are compared with original Matsumoto and Sigrist calculations[13]. It is seen in Fig. 3d that the maximal field produced by a current of 0.6 μA is 0.56 mG. Essentially, if the chiral surface and Meissner currents obtained in the original Matsumoto-Sigrist calculation are reduced by a numerical factor without altering their distribution, the upper bound on the magnetic field created by the chiral surface and Meissner currents set by the scanning SQUID studies is roughly two orders of magnitude smaller than that predicted by the Matsumoto and Sigrist, corresponding to a magnetic field of tens of mG. The current experiment has shown that a reduction factor of $10^3$ to $10^4$ is needed to account for the experimental results presented above.

It is reasonable to ask whether the $R_J$ *vs*. $I_{cry}$ measurements on Junction A would be affected by the residue magnetic field from the Earth field. In all three separate cooling downs with the superconducting magnet in open circuit. The $R_J$ *vs*. $I_{cry}$ measurements were performed before any current is flown into the superconducting magnet. Because the sample sit in the center of the solenoid of the superconducting magnet with a bore size of 3.8 cm and a length of 25.4 cm, the earth field trapped in the magnet, which became a constant after the superconducting magnet is cooled down to roughly 15 K, will indeed be present and along the direction of the axis of the magnet, the junction, and the *ab* plane of the crystal in the normal state of the sample. However, after the samples became superconducting below roughly 1 K, the tiny magnetic flux from the Earth field trapped in the superconducting magnet will be excluded by the Al and $Sr_2RuO_4$ crystal due to the Meissner effect, which makes the local residual field on the junction plane tiny considering that the Al film is thick (200 nm), the crystal of $Sr_2RuO_4$ is of the macroscopic size, and the junction barrier is so thin in comparison with both superconducting electrodes.

A natural question is why the offset current of $I_{0,cry}$ is of the same sign measured in different cooling downs. As pointed out above, the quantum oscillation pattern we observed in Sample A suggests that a single domain is probed. It was found theoretically that the direction of the chiral edge current depends on the surface orientation due to the subtleties in the electronic band structure of $Sr_2RuO_4$[64], as shown schematically in Fig. 4a. Given that the point at which the chiral edge current changes the direction must correspond to a domain wall, the likely scenario is that the chiral surface current was seen to feature a fixed sign in Junction A because of the specific crystalline orientation of the plane of the junction, which cannot be changed because of the cooling downs. In fact, the observation of a single-signed $I_{0,cry}$ is fully consistent with the single-domain quantum oscillations described above.

It is highly desirable that a control experiment on what is presented above be performed. In this regard, it is important to note that the Meissner current generated by ***H*** (applied in the *ab* plane of the crystal) is perpendicular to the chiral surface current. As a result, the chiral surface current will not be canceled by the variation of the applied field. However, placing a permanent magnet (which produces a field of a few hundred Gauss near the magnet) outside the cryostat (about a meter away from the dewar) could produce a magnetic field with a component perpendicular to the *ab* plane on the crystal surface and a Meissner current in the direction of chiral surface current. Even though its precise value is not known, the Meissner current generated by the permanent magnet on the surface is expected to be small because not only is the magnet placed far away from the junction, but the field is also greatly reduced by the shielding of the large superconducting solenoid, making the component of the residual field along the *c* axis tiny. Importantly, however, the sign of this component will be reversed if the magnet is flipped in its orientation (see Insets of Figs. 4c and d). Our control experiment with and without the permanent magnet including flipping of the magnet, yielded the expected responses in $R_J$ *vs.* $I_{cry}$ and values of $I_{0,cry}$ shown in Figs. 4b-d. It is interesting to note that the local field produced on the junction plane by the permanent magnet placed away from the cryostat (inferred from the conversion of the $I_{0,cry}$ value to local field in the London approximation presented above) is tiny. However, this is consistent with the effect of the Meissner screening by the sample discussed above. The values of $I_{0,cry}$ obtained in the control experiment are comparable with those shown in Figs. 3b and c, supporting the validity of using $I_{0,cry}$ to estimate the size of the magnetic flux produced by the chiral surface and Meissner currents.

The physical origin of the discrepancy between the experimental and theoretical results from Matsumoto and Sigrist[13] on the chiral edge current in $Sr_2RuO_4$ is complex. The magnetic flux generated by both chiral surface and Meissner currents is confined in a narrow spatial region where the two currents flow. In the theoretical calculations of the net magnetic flux, the crystal is assumed to occupy half of the infinite space. Experimentally, for both the scanning SQUID[41,42] and Hall probe[43] measurements, the field sensor is placed at least 1 μm above the finite-size crystal, which at the best can be seen as occupying a quarter of the infinite space. The consequences of these two configurations have not been examined. In addition, the possible presence of domains and domain walls in these samples may modify the current configuration and could significantly reduce the magnetic flux signal detected above the crystal. Theoretically, the disagreement between the experiments and the theory was examined in detail[65]. Various theoretical proposals in which the chiral surface current may be reduced from that of Matsumoto and Sigrist by orders of magnitude have been put forward[66,67,68,69,70,71,72,73,74,75]. In addition, to detect nonreciprocal signal in the Josephson effect in $Sr_2RuO_4$/Nb junctions[44], which is directly related to the magnetic flux generated by the chiral surface and Meissner currents[45,76], the induced magnetic field estimated from the current work (which is consistent with the upper bound set by the previous work using the scanning SQUID) is much smaller than the detection limit in Ref. 44. As a result, even though the signal for the nonreciprocal effect under current-filed inversion test was not detected, it should not serve as evidence against TRS breaking in $Sr_2RuO_4$.

In conclusion, we successfully prepared high-quality ramp Josephson junctions of Al/Ti/$Sr_2RuO_4$ by conventional and unconventional nanofabrication techniques. This work helps lay the foundation for the full determination of the angle dependence of the superconducting OP in $Sr_2RuO_4$ by phase-sensitive measurements and a final resolution of the current spin-singlet *vs*. triplet debate on this material. We also developed a new approach to detect the chiral surface current in $Sr_2RuO_4$ based on the sensitive dependence of the junction resistance on the supercurrent flowing in the crystal of $Sr_2RuO_4$ on which the Josephson junction was prepared. The magnetic field produced by the chiral edge and the associated Meissner currents were found to be $10^3$ to $10^4$ times smaller than that obtained by Matsumoto and Sigrist. Our work is therefore a positive step forward towards building a paradigm for unconventional superconductivity relevant for quantum technologies.

**Acknowledgements.**

We would like to thank A. J. Leggett, W. Huang, M. Sigrist, C. Kallin, Y. Maeno, S-B. Chung, V. Vakaryuk, S-K. Yip, and J. Kirtley for useful discussions. Research at Penn State was supported by the National Science Foundation (NSF) under Grant No. DMR-2312899. Crystal growth was further supported by the Penn State 2D Crystal Consortium–Materials Innovation Platform (2DCC-MIP) under NSF Cooperative Agreement DMR-2039351. The TEM images were obtained at the TEM facility at FSU, supported by the Florida State University Research Foundation, NSF-DMR-1157490, and the State of Florida.

## Figures and Figure Captions

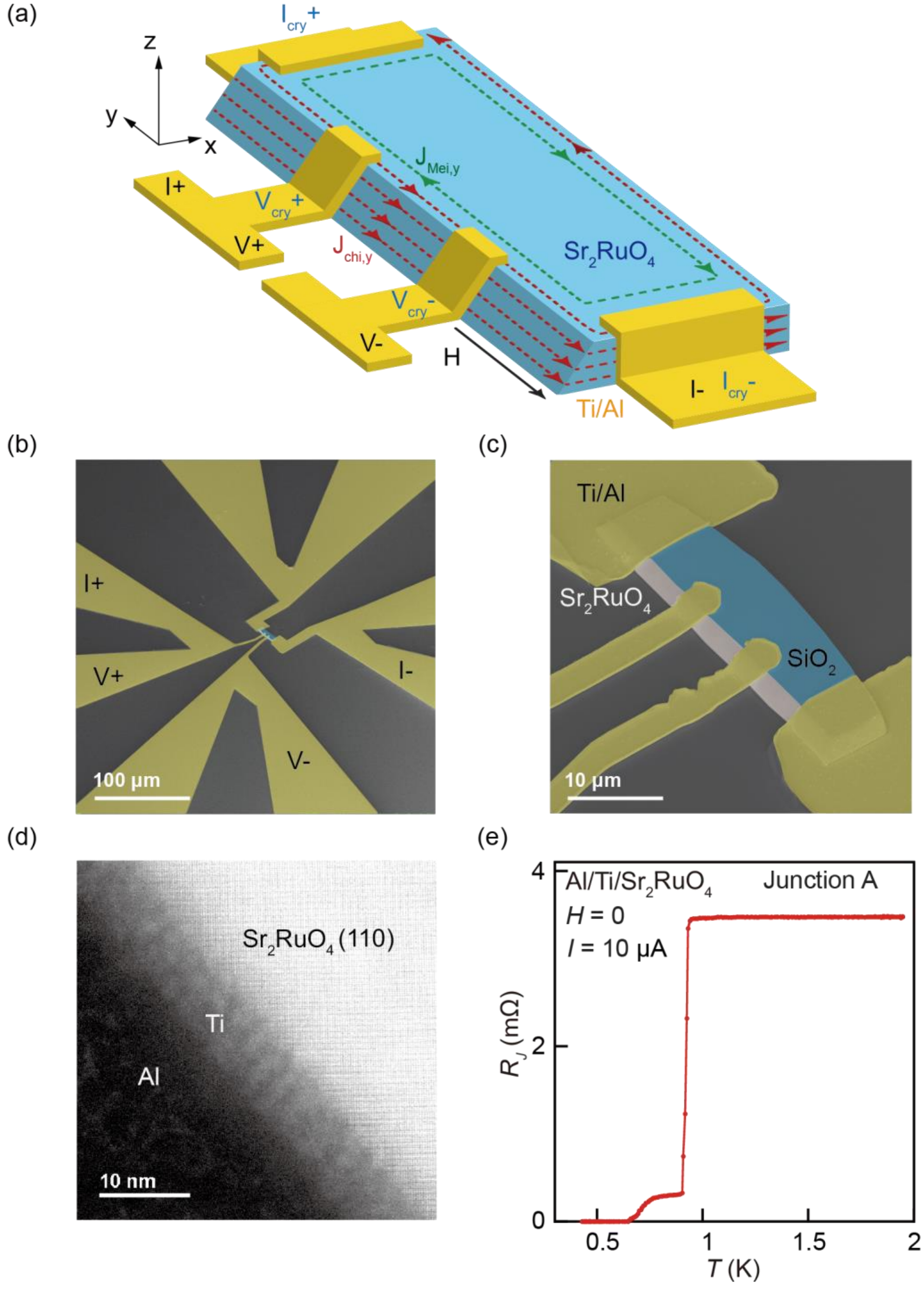

**FIG. 1**. (a) Schematic of the Al/Ti/$Sr_2RuO_4$ ramp Josephson junction prepared on a thin crystal of $Sr_2RuO_4$ obtained by mechanical exfoliation. A supercurrent of $I_{cry}$ flowing through the crystal is supplied from the ends of the crystal. The Meissner current $J_{Mei,y}$ (shown in green) responding to the presence of the chiral surface current flows in the opposite direction to $J_{chi,y}$ (shown in red) towards the interior of the crystal, both of which are along the y axis in the coordinate system shown in the schematic. The ramp was prepared using the focused ion beam (FIB) with the surface after the FIB cut cleaned by low-energy ion milling. The magnetic field, ***H***, is applied along the *ab* plane of the crystal as well as the junction plane. The electrical leads used to measure the crystal ($I_{cry}$+, $I_{cry}$-, $V_{cry}$+, $V_{cry}$-) and junction (I+, I-, V+, V-) are labeled. The insulating layer of $SiO_2$ deposited on top of the $Sr_2RuO_4$ crystal is not shown. (b-c) Scanning electron microscopy (SEM) image of the sample at two magnifications. The width the Al/Ti leads is 3 μm. The thicknesses of the Ti and Al layers are 5 nm and 200 nm, respectively. The crystal plate of $Sr_2RuO_4$ is around 1 - 2 μm thick. (d) High angle annular dark field scanning transmission electron microscopy (STEM) image of cross-sectional view of an Al/Ti/$Sr_2RuO_4$ junction. The ramp is 50 degrees from the *ab*-plane. (e) Junction resistance as a function of temperature, $R_J$ (T), for Junction A measured at $I$ = 10 μA, showing zero junction resistance below 0.65 K. The critical temperature for Al is around 0.9 K.

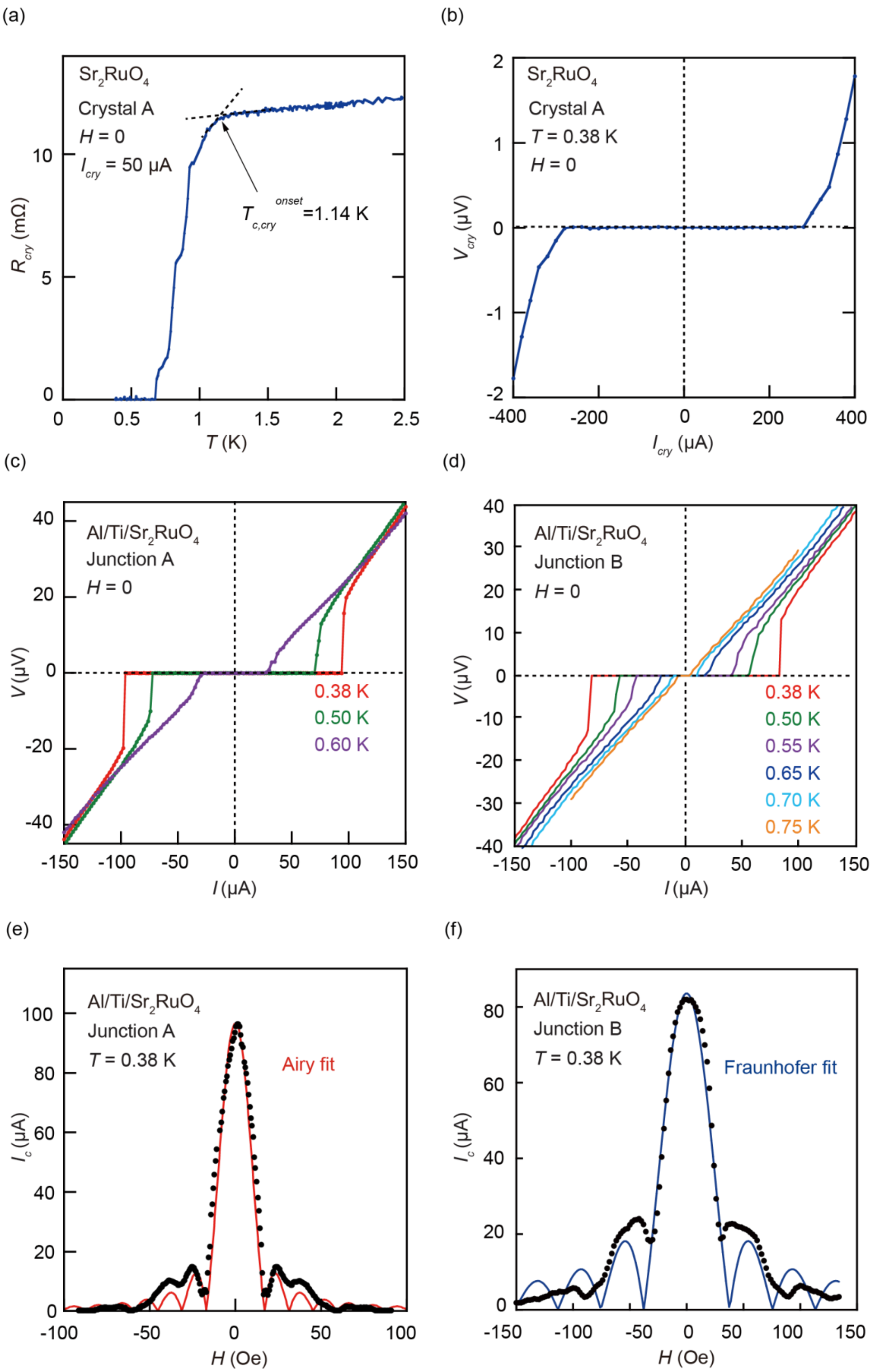

**FIG. 2**. (a) Crystal resistance as a function of temperature, $R_{cry}$ (T), of Crystal A (the crystal of $Sr_2RuO_4$ used for Junction A) measured at $I_{cry}$ = 50 μA. The onset and zero-resistance superconducting transition temperatures are 1.14 and 0.68 K, respectively. (b) Voltage-current ($V_{cry} - I_{cry}$) curve of Crystal A at 0.38 K, showing a critical current $I_{c,cry}$ of 280 μA. (c-d) Zero-field $V - I$ curves of Junction A at 0.38 K, 0.5 K and 0.6 K, Junction B at 0.38 K, 0.5 K, 0.55 K, 0.65 K, 0.7 K and 0.75 K. The Josephson coupling signaled by the zero voltage is clearly seen. (e-f) Dependence of junction critical current $I_c$ on applied magnetic field $H$ of Junctions A and B measured at 0.38 K. Solid red line represents Airy fit while solid blue line represents Fraunhofer fit to the data. A comparison of the two fits is provided in the Supplementary Materials (SM). The fitting of our data provides strong evidence for single-domain Josephson coupling between Al and $Sr_2RuO_4$.

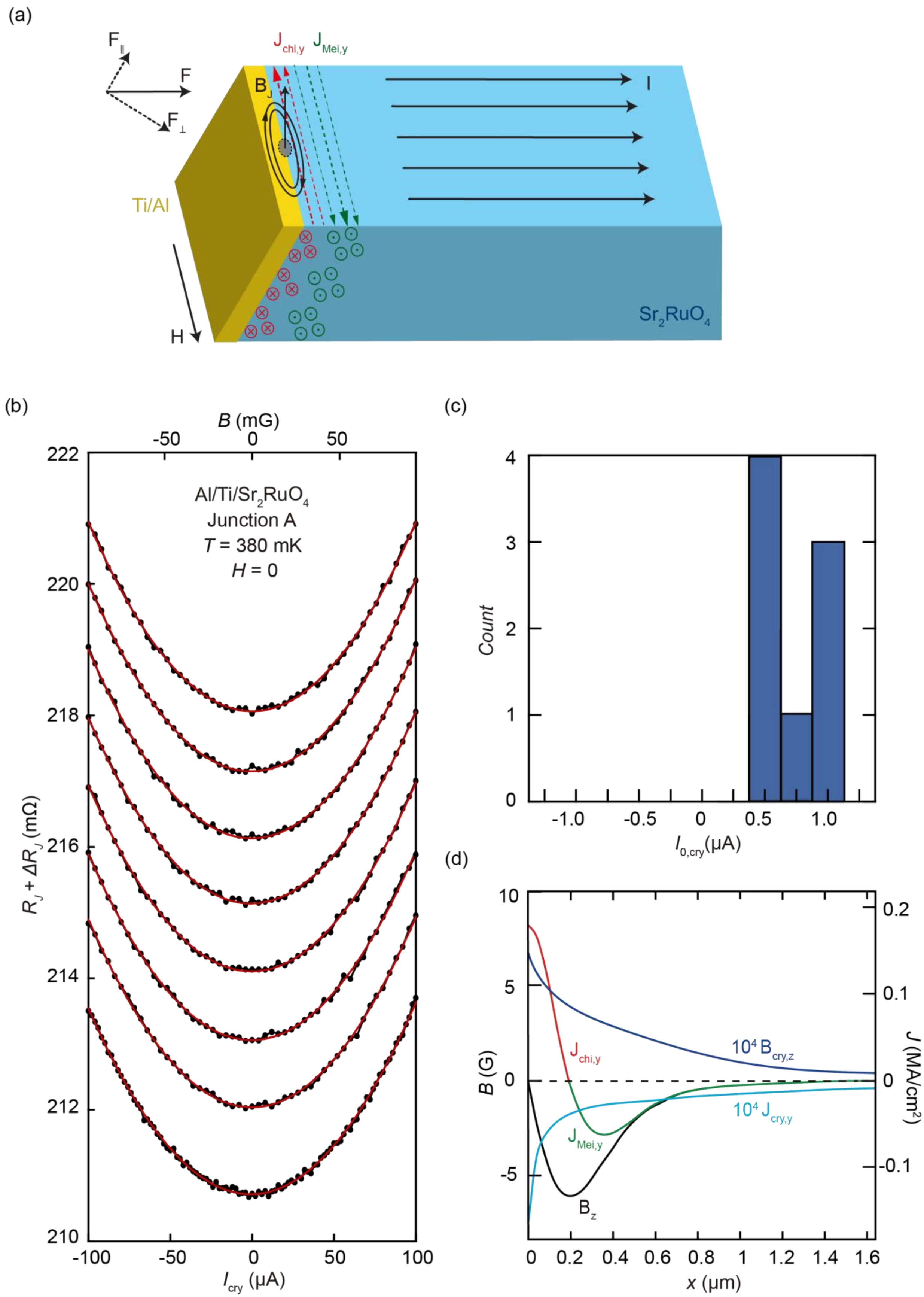

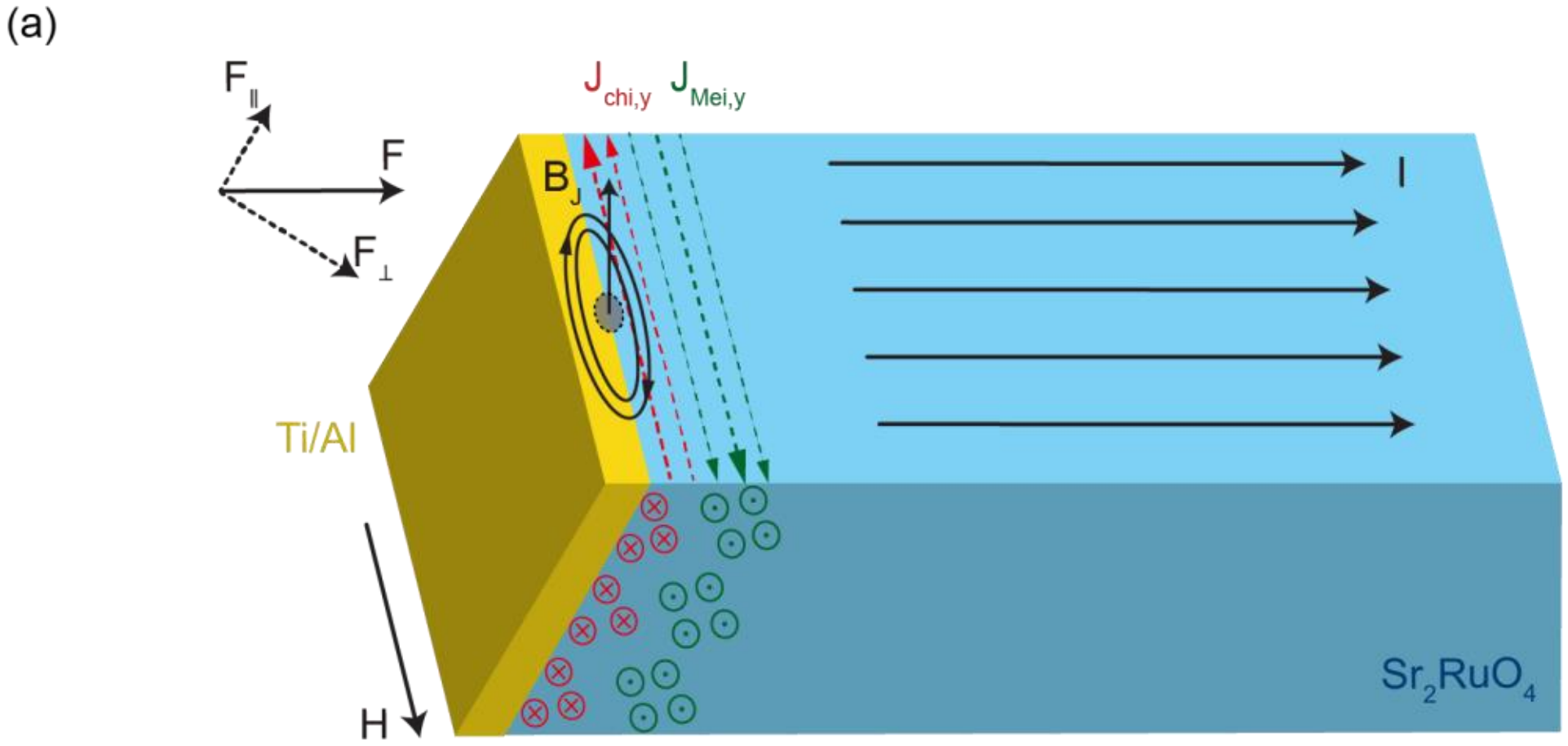

**FIG. 3**. (a) Schematic of a Josephson vortex in the Al/Ti/$Sr_2RuO_4$ ramp Josephson junction with the magnetic flux line along the *c* axis. (b) $R_J$ measured at a junction current of $I = 100$ μA at 0.38 K with the supercurrent flowing through the crystal, $I_{cry}$, varied. Eight separate measurements were carried out in zero magnetic fields. The junction was cooled down from the normal state to 0.38 K three times. Except for the bottom curve, all others are vertically offset by 1 mΩ for clarity. The red curves are parabolic fits for each set of data. (c) Histogram of the values of $I_{0,cry}$ obtained from fitting in Fig. 3b. The average is 0.6 μA. (d) Distribution of $J_{chi,y}$ (x), $J_{Mei,y}$ (x) and corresponding field $B_z$ (x), as calculated by Matsumoto and Sigrist[13] along with distribution $J_{cry,y}$ (x) and $B_{cry,z}$ (x) calculated in the London approximation with the actual dimensions of Crystal A. For presentation purposes, values of $J_{cry,y}$ (x) and $B_{cry,z}$ (x) are scaled by $10^4$ in the plot.

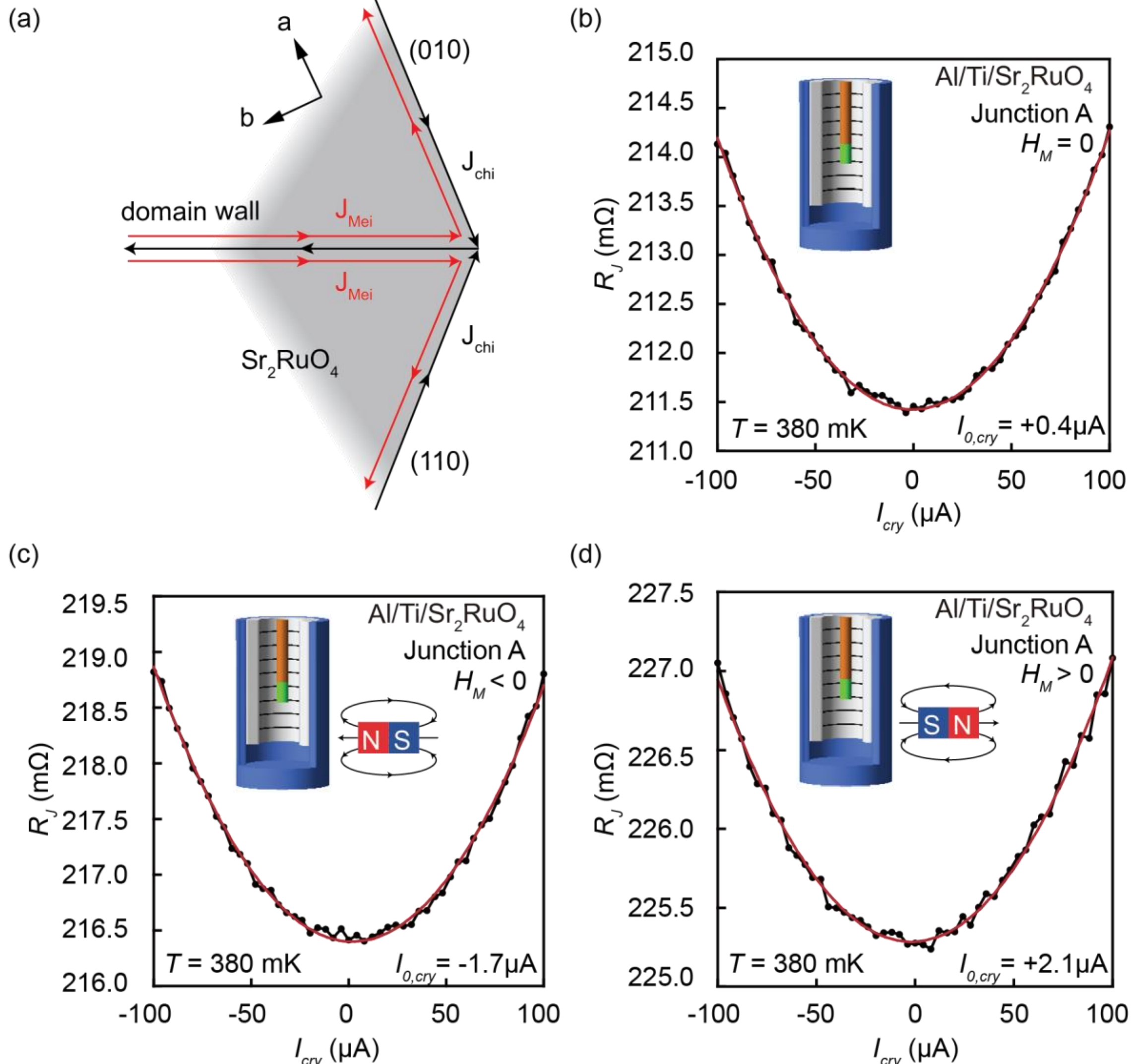

**FIG. 4.** (a) Schematic showing that the chiral surface current in $Sr_2RuO_4$ depends on the surface orientation. The two crystal surface orientations are indicated. The domain wall will form where the opposing chiral edge currents meet. (b-d) A control experiment using a permanent magnet placed outside the cryostat, which generated a magnetic field $H_M$ on the crystal surface with a component along the same direction as that produced by the chiral surface and Meissner currents. $R_J$ measured at a junction current of $I = 100$ μA as a function of $I_{cry}$, with and without the presence of the magnet, at $T = 0.38$ K, including curves taken for opposite magnet orientations. The red curves are parabolic fits. The insets show the schematics of the experimental setup. A sign change in $I_{0,cry}$ is seen after flipping the orientation of the permanent magnet.

[18] R. Jin, Y. Liu, Z. Q. Mao, and Y. Maeno, "Experimental observation of the selection rule in Josephson coupling between In and $Sr_2RuO_4$," *Europhys. Lett*. **51**, 341 (2000).

[19] I. Žutić and I. Mazin, "Phase-sensitive tests of the pairing state symmetry in $Sr_2RuO_4$," *Phys. Rev. Lett*. **95**, 217004 (2005).

[20] V. Ambegaokar and A. Baratoff, "Tunneling between superconductors," *Phys. Rev. Lett*. **10**, 486 (1963). erratum. *Phys. Rev. Lett*. **11**, 104 (1963).

[21] K. D. Nelson, Z. Q. Mao, Y. Maeno, and Y. Liu, "Odd-parity superconductivity in $Sr_2RuO_4$," *Science* **306**, 5699 (2004).

[22] K. Ishida, H. Mukuda, Y. Kitaoka, K. Asayama, Z. Q. Mao, Y. Mori, and Y. Maeno. "Spin-triplet superconductivity in Sr2RuO4 identified by $^{17}$O Knight shift." *Nature* **396**, 658-660 (1998).

[23] A. Pustogow, Y. Luo, A. Chronister, Y.-S. Su, D. A. Sokolov, F. Jerzembeck, A. P. Mackenzie, C. W. Hicks, N. Kikugawa, S. Raghu, E. D. Bauer, and S. E. Brown, "Constraints on the superconducting order parameter in $Sr_2RuO_4$ from oxygen-17 nuclear magnetic resonance," *Nature* **574**, 7776 (2019).

[24] K. Ishida, M. Manago, K. Kinjo, and Y. Maeno, "Reduction of the $^{17}$O Knight Shift in the Superconducting State and the Heat-up Effect by NMR Pulses on $Sr_2RuO_4$," *J. Phys. Soc. Jpn*. **89**, 034712 (2020).

[25] A. Chronister, A. Pustogow, N. Kikugawa, D. A. Sokolov, F. Jerzembeck, C. W. Hicks, A. P. Mackenzie, E. D. Bauer, and S. E. Brown, "Evidence for even parity unconventional superconductivity in $Sr_2RuO_4$," *Proc. Nat. Acad. Sci.* **118**, e2025313118 (2021).

[26] A. N. Petsch, M. Zhu, M. Enderle, Z. Q. Mao, Y. Maeno, I. I. Mazin, and S. M. Hayden, "Reduction of the spin susceptibility in the superconducting state of $Sr_2RuO_4$ observed by polarized neutron scattering," *Phys. Rev. Lett*. **125**, 217004 (2020).

[27] A. J. Leggett, "Spin susceptibility of a superfluid fermi liquid," *Phys. Rev. Lett*. **14**, 536 (1965).

[28] A. J. Leggett and Y. Liu, "Symmetry properties of superconducting order parameter in $Sr_2RuO_4$," *J. Supercond. Nov. Magn*. **34**, 1647 (2021).

[29] A. J. Leggett, "Quantum Liquids: Bose condensation and Cooper pairing in condensed-matter systems," Oxford University Press, Oxford (2006).

[30] K. Seiler, C. Gros, T. M. Rice, K. Ueda, and Dieter Vollhardt. "Crossover from Fermi liquid to classical behavior of normal $^3$He in the model of almost localized fermions," *J. Low Temp. Phys*. **64**, 195-221 (1986).

[31] D. N. Paulson, R. T. Johnson, and J. C. Wheatley, "Static Nuclear Magnetism in Extraordinary Liquid He-3," *Phys. Rev. Lett.* **31**, 746 (1973).

[32] Yoshiteru Maeno, Koji Yoshida, Hiroaki Hashimoto, Shuji Nishizaki, Shin-ichi Ikeda, Minoru Nohara, Toshizo Fujita, Andrew P. Mackenzie, Nigel E. Hussey, J. Georg Bednorz, and Frank

[75] H.-T. Liu, W. Chen, J.-X. Yin, C. Liu, and W. Huang, "Examining the possibility of chiral superconductivity in $Sr_2RuO_4$ and other compounds via applied supercurrent," *Phys. Rev. B* **109**, 014508 (2024).

[76] B. Zinkl, K. Hamamoto and M. Sigrist, Symmetry conditions for the superconducting diode effect in chiral superconductors. *Phys. Rev. Res.* **4**, 033167 (2022).